\documentstyle[11pt]{article}
\begin{document}
\newtheorem{Theorem}{Theorem}
\newtheorem{Lemma}{Lemma}

\newcommand{\qed}{\hbox{\rule[-2pt]{3pt}{6pt}}}

\title{Convergence of simulated annealing\\
by the generalized transition probability
}
\author{Hidetoshi Nishimori and Jun-ichi Inoue\\ \\
Department of Physics, Tokyo Institute of Technology\\
 Oh-okayama, Meguro-ku, Tokyo 152, Japan
}

\date{\today}

\maketitle

\begin{abstract}
We prove weak ergodicity of the inhomogeneous
Markov process generated by the generalized transition 
probability of Tsallis and Stariolo under power-law decay
of the temperature.
We thus have a mathematical foundation to conjecture
convergence of simulated annealing
processes with the generalized transition probability
to the minimum of the cost function.
An explicitly solvable example in one dimension
is analyzed in which
the generalized transition probability leads to a fast convergence
of the cost function to the optimal value.
We also investigate how far our arguments depend upon the
specific form of the generalized transition probability
proposed by Tsallis and Stariolo.
It is shown that a few requirements
on analyticity of the transition probability are sufficient
to assure fast convergence in the case of the solvable
model in one dimension.

\end{abstract}
%%%%%%%%%%%%%%%%%%%%%%%%%%%%%%%%%%%%%%%%%%%%%%%%%%%%%%%%%%
\section{Introduction}
%%%%%%%%%%%%%%%%%%%%%%%%%%%%%%%%%%%%%%%%%%%%%%%%%%%%%%%%%%
Simulated annealing has been a powerful tool for combinatorial
optimization problems \cite{Kirkpatrick,Uesaka,Aarts}.
To find the minimum of a cost function, one introduces a
stochastic process similar to Monte Carlo simulations
in statistical mechanics with a control parameter
corresponding to the temperature to allow the system
to escape from local minima.
By gradually decreasing the temperature one searches for
increasingly narrower regions in the phase space
closer to the optimal state, eventually reaching the
optimal state itself in the infinite-time limit.

A very important factor in such processes is the
annealing schedule, or the rate of decrease of temperature.
If one lowers the temperature too quickly, the system
may end up in one of the local minima.
On the other hand, a very slow decrease of temperature
would surely bring the system to
the true minimum.  However, such a slow process is not
practically useful.
One therefore has to determine carefully how fast to
decrease the temperature in simulated annealings.
On this problem,
Geman and Geman \cite{Geman} proved
that the decrease of temperature as
$T= {\rm const}/\log t$,
with the proportionality constant
roughly of the order of the system size,
guarantees convergence
to the optimal state for a wide class of combinatorial
optimization problems.
This inverse-log law is still too slow for most
practical purposes.
Nevertheless this result serves as a mathematical
background of empirical investigations by numerical methods.

There have been a few proposals to accelerate the
annealing schedule by modifying the transition probabilities
used in the conventional simulated annealing.
Szu and Hartly \cite{Szu} pointed out 
for a problem defined in a continuum space
that occasional non-local samplings significantly
improve the performance, leading to an annealing schedule
inversely proportional to time $T={\rm const}/t$.
This non-local sampling corresponds to
modification of the generation probability (or, more precisely,
the neighbourhood) to be defined later.
Tsallis and Stariolo \cite{Tsallis} proposed to modify
the acceptance probability generalizing the
usual Boltzmann form in addition to the generation
probability (which they call the visiting distribution).
Numerical investigations
show faster convergence to the optimal state by annealing
processes using their generalized transition probability or
its modifications \cite{Tsallis,Penna,Andre}.
Szu and Hartley and Tsallis and Stariolo proved that the
modified {\it generation} probability assures convergence
to the optimal state under a power-low decrease of
the temperature as a function of time.
However there has been no mathematically rigorous
argument on the convergence under the generalized
{\it acceptance} probability of Tsallis and Stariolo.

We prove in the present paper that the inhomogeneous
Markov process generated 
by the generalized transition (acceptance) probability of
Tsallis and Stariolo satisfies the property of weak
ergodicity under an annealing schedule inversely proportional
to the power of time.
Rigorously speaking, weak ergodicity (which roughly means asymptotic
independence of the probability distribution from the initial
condition) itself does not immediately guarantee the convergence
to the optimal state.
Nevertheless our result is expected to be close enough to this final goal
because the probability distribution would depend upon the
initial condition if the annealing schedule is not appropriately chosen.

Various definitions are given in the next section.
The proof of our main theorem appears in section 3.
An example of fast convergence by the generalized transition probability
is discussed in section 4 for a parameter range not covered by the
theorem in section 3.
In section 5 we investigate if we may further generalize the transition 
probability in the case of the simple model discussed in section 4.
The final section is devoted to discussions
on the significance of our result.

%%
%%
%%
%%
%%%%%%%%%%%%%%%%%%%%%%%%%%%%%%%%%%%%%%%%%%%%%%%%%%%%%%%
\section{Inhomogeneous Markov chain}
%%%%%%%%%%%%%%%%%%%%%%%%%%%%%%%%%%%%%%%%%%%%%%%%%%%%%%%
%
Let us first list up various definitions to fix notations.
We consider a problem of combinatorial optimization with the
space of states denoted by $\cal S$.
The size of $\cal S$ is finite.
The {\it cost function} $E$ is a real single-valued function on
${\cal S}$.
The goal of a combinatorial optimization problem is to find
the minimum (or minima) of the cost function.
For this purpose we introduce the process of simulated annealing
using the Markov chain generated by the {\it transition probability}
from state $x (\in {\cal S})$ to state $y (\in {\cal S})$ at time step $t$:
\begin{equation}
  G(x, y; t)= \left\{
    \begin{array}{ll}
      P(x, y) A(x, y; T(t)) & \quad (x\ne y) \\
      1-\sum_{z(\ne x)}  P(x, z) A(x, z; T(t)) & \quad (x=y)
    \end{array}
    \right.
    ,
    \label{trp}
 \end{equation}
where $P(x, y)$ is the {\it generation probability}
\begin{equation}
   P(x, y)\left\{
    \begin{array}{ll}
      >0 & \quad (y\in {\cal S}_x) \\
      = 0 & \quad (\mbox{otherwise})
    \end{array}
    \right.
\end{equation}
with ${\cal S}_x$ the {\it neighbourhood} of $x$
(the set of states that can be reached by a single step
from $x$), and
$A(x, y; T)$ is the {\it acceptance probability}.
In the case of the generalized transition probability,
the acceptance probability is given as \cite{Tsallis}
\begin{eqnarray}
  A(x, y; T) &=& \min \{1, u(x, y; T)\} \nonumber\\
  u(x, y; T)&=& \left( 1+(q-1)\frac{E(y)-E(x)}{T} \right)
    ^{1/(1-q)}
  \label{acceptance}
\end{eqnarray}
where $q$ is a real parameter.
For technical reasons we have to restrict ourselves to the
region $q>1$ in this and the next sections.
This acceptance probability reduces to the
usual Boltzmann form in the limit $q\to 1$.
The present Markov chain is {\it inhomogeneous}, i.e.,
the transition probability (\ref{trp}) depends on the
time step $t$ through time dependence of $T(t)$.

We choose the {\it annealing schedule}, or the $t$-dependence of the
parameter $T$ (the {\it temperature}), as
\begin{equation}
T(t)=\frac{b}{(t+2)^c}  \quad (b, c>0, t=0,1,2,\cdots) .
\label{schedule}
\end{equation}

To analyze the inhomogeneous Markov chain generated by the above
transition probability, we introduce the {\it transition matrix}
$G(t)$ with the element
\begin{equation}
 [G(t)]_{x,y}=G(x, y; t) .
\end{equation}
Let us write the set of probability distributions on $\cal S$
as $\cal P$.
A probability distribution $p (\in {\cal P})$ may be regarded
as a vector with the component
$[p]_x=p(x) (x\in {\cal S})$.
Using this matrix-vector notation,
the probability distribution at time step $t$, starting from
an initial distribution $p_0 (\in {\cal P})$ at time $s$,
is written as
\begin{equation}
  p(s, t)=p_0 G(s, t)=p_0 G(s)G(s+1)\cdots G(t-1) .
  \label{Gst}
\end{equation}
The {\it coefficient of ergodicity} is defined as
\begin{equation}
  \alpha (G)=1-\min \{ \sum_{z\in {\cal S}} \min \{
   G(x, z), G(y, z)\} | x, y\in {\cal S}\} .
\end{equation}

We shall prove in the next section the property of
{\it weak ergodicity} for the present Markov chain,
which means that the probability distribution function after
sufficiently long time becomes independent of the initial condition:
\begin{equation}
  \forall s\ge 0 : \lim_{t\to\infty}\sup
    \{ \|p_1(s, t)-p_2(s, t)\|  \mid p_{01}, p_{02}\in {\cal P}\}=0
\label{werg}
\end{equation}
where $p_1(s, t)$ and $p_2(s, t)$ are the probability distributions
with different initial conditions $p_{01}$ and $p_{02}$:
\begin{eqnarray}
  p_1(s, t)&=&p_{01} G(s, t) \\
  p_2(s, t)&=&p_{02} G(s, t) .
\end{eqnarray}
The norm is defined by
\begin{equation}
  \|p_1-p_2\| = \sum_{x\in {\cal S}} |p_1(x)-p_2(x)| .
\end{equation}
Although we focus our attention on weak ergodicity in the present paper,
it may be useful as a reference to recall the
definition of {\it strong ergodicity}:
\begin{equation}
  \exists r\in {\cal P}, \forall s\ge 0 :
    \lim_{t\to \infty}\sup\{ \| p(s, t)-r\| \mid p_0\in {\cal P}\}=0 .
\end{equation}

The following theorems give criteria for weak and strong ergodicity
\cite{Uesaka,Aarts}:
\begin{Theorem}[Condition for weak ergodicity]\label{weak}
An inhomogeneous Markov chain is weakly ergodic if and only if
there exists a strictly increasing sequence of positive numbers
\[
   t_0 < t_1 < \cdots <t_i<t_{i+1}<\cdots
\]
such that
\begin{equation}
  \sum_{i=0}^\infty (1-\alpha (G(t_i, t_{i+1})))=\infty .
  \label{condweak}
\end{equation}
\end{Theorem}
\begin{Theorem}[Condition for strong ergodicity]\label{strong}
An inhomogeneous Markov chain is strongly ergodic if it satisfies
the following conditions:
\begin{enumerate}
   \item  it is weakly ergodic
   \item there exists $p_t\in {\cal P} (\forall t\ge 0)$ such that
      $p_t=p_tG(t)$
   \item $p_t$ satisfies
     \begin{equation}
       \sum_{t=0}^\infty \| p_t-p_{t+1} \| < \infty .
         \label{scond}
      \end{equation}
\end{enumerate}
\end{Theorem}
%
%
%%%%%%%%%%%%%%%%%%%%%%%%%%%%%%%%%%%%%%%%%%%%%%%%%%%%%%%%%%%%
\section{Weak ergodicity}
%%%%%%%%%%%%%%%%%%%%%%%%%%%%%%%%%%%%%%%%%%%%%%%%%%%%%%%%%%%%
We prove in the present section that the condition in Theorem \ref{weak}
is satisfied by the present inhomogeneous Markov
chain generated by the generalized transition probability.
The argument closely follows that for the conventional
Boltzmann-type transition probability \cite{Geman,Uesaka,Aarts}.
We need the following Lemma for this purpose.
\begin{Lemma}[Lower bound on the transition probability]
The elements of the transition matrix satisfy the following bounds.
For off-diagonal elements,
\begin{equation}
  P(x, y)>0 \Rightarrow \forall t\ge 0 :
    G(x, y; t) \ge w \left( 1+\frac{(q-1)L}{T(t)}\right)^{1/(1-q)} ,
      \label{bound1}
\end{equation}
and for diagonal elements,
\begin{equation}
  \forall x\in {\cal S}-{\cal S}_m, \exists t_1>0, \forall t\ge t_1 :
  G(x, x; t) \ge w \left( 1+\frac{(q-1)L}{T(t)}\right)^{1/(1-q)} 
      \label{bound2}
\end{equation}
where ${\cal S}_m$ is the set of locally maximum states
\begin{equation}
  {\cal S}_m =\{ x | x\in {\cal S}, \forall y \in {\cal S}_x : E(y)\le E(x)\} ,
\end{equation}
$L$ denotes the maximum change of the cost function by a single step
\begin{equation}
  L=\max \{ | E(x)-E(y) | \mid P(x, y)>0\}
\end{equation}
and $w$ is the minimum value of $P(x, y)$
\begin{equation}
  w=\min \{ P(x, y) \mid P(x, y)>0, x, y\in {\cal S} \} .
\end{equation}
\end{Lemma}
%%%%%%%%%%%%%%%%%%%%%%%%%%%%%%%%%%%%%%
{\it Proof}.

First we prove (\ref{bound1}). When $E(y)-E(x)>0$, we have
$u(x, y; T(t))\le 1$ and thus
\begin{eqnarray}
   G(x, y; t) &=& P(x, y)A(x, y; T(t))
   \nonumber\\
   &\ge & w\,  \min\{1, u(x, y; T(t))\}
      \nonumber\\
      &=& w\, u(x, y; T(t))
         \nonumber\\
      &\ge & w \left( 1+\frac{(q-1)L}{T(t)}\right)^{1/(1-q)} .
\end{eqnarray}
If $E(y)-E(x)\le 0$,  $u(x, y; T(t))\ge 1$ and therefore
\begin{eqnarray}
   G(x, y; t) &\ge & w \,  \min\{ 1, u(x, y; T(t))\}
      \nonumber\\
      &=& w \nonumber\\
      &\ge &w \left( 1+\frac{(q-1)L}{T(t)}\right)^{1/(1-q)} .
\end{eqnarray}

We next prove (\ref{bound2}).
Since $x\in {\cal S}-{\cal S}_m$,
there exists a state $y\in {\cal S}_x$ satisfying $E(y)-E(x) > 0$.
For such a state $y$,
\begin{equation}
  \lim_{t\to\infty} u(x, y; T(t)) =0
\end{equation}
and consequently
\begin{equation}
  \lim_{t\to\infty} \min\{ 1,  u(x, y; T(t))\} =0 .
\end{equation}
Then $\min\{ 1,  u(x, y; T(t))\}$ can be made arbitrarily small for
sufficiently large $t$.  More precisely, there exist
$t_1>0$ and $0<\epsilon <1$ such that
\begin{equation}
  \forall t\ge t_1 : \min\{ 1, u(x, y; T(t))\} < \epsilon .
\end{equation}
We therefore have
\begin{eqnarray}
  \sum_{z\in {\cal S}}P(x, z) A(x, z; T(t))
    &=& P(x, y) \min\{ 1, u(x, y; T(t))\}
        +\sum_{z\in {\cal S}-\{y\} } P(x, z) \min\{ 1, u(x, z; T(t))\}
     \nonumber\\
  &<& P(x, y)\epsilon +\sum_{z\in {\cal S}-\{y\} } P(x, z)
     \nonumber\\
   &=& -(1-\epsilon )P(x, y) +1 .
\end{eqnarray}
The diagonal element of (\ref{trp}) thus satisfies
\begin{eqnarray}
  G(x, x; t) &\ge &(1-\epsilon ) P(x, y) \nonumber \\
  &\ge & w\left( 1+\frac{(q-1)L}{T(t)}\right)^{1/(1-q)}
\end{eqnarray}
where we have used that the last factor can be
made arbitrarily small for sufficiently large $t$.
\qed
\vspace{2mm}

%%%%%%%%%%%%%%%%%%%%%%%
We use the following notations in the proof of weak ergodicity.
The minimum number of state transitions to reach $y$ from $x$
(or {\it vice versa}) is written as $d(x, y)$.
One can then reach any state from $x$ within $k(x)$ steps:
\begin{equation}
  k(x) =\max\{ d(x, y) | y\in {\cal S}\} .
\end{equation}
The minimum of $k(x)$ for $x\in {\cal S}-{\cal S}_m$
is denoted as $R$, and the state giving this
minimum value is $x^*$:
\begin{eqnarray}
    R&=&\min\{ k(x) | x\in {\cal S}-{\cal S}_m \}
       \label{R}
       \\
    x^* &=& {\rm arg}\min\{ k(x) | x\in {\cal S}-{\cal S}_m \} .
\end{eqnarray}
%
%%%%
\begin{Theorem}[Weak ergodicity]\label{convergence}
The inhomogeneous Markov chain defined in section 2 is
weakly ergodic if $0<c\le (q-1)/R$.
\end{Theorem}
{\it Proof}.

Consider a transition from state $x$ to $x^*$.  According
to the definition (\ref{Gst}) of the double-time transition matrix,
we have
\begin{equation}
 G(x, x^*; t-R, t)=
 \sum_{x_1,\cdots ,x_{R-1}} G(x, x_1; t-R)G(x_1, x_2; t-R+1)
    \cdots G(x_{R-1}, x^*; t-1) .
       \label{Gxx}
\end{equation}
From the definitions of $x^*$ and $R$, there exists at least
one sequence of transitions to reach $x^*$ from $x$ within
$R$ steps such that
\begin{equation}
 x\ne x_1 \ne x_2\ne \cdots \ne x_k =x_{k+1}\cdots =x_R =x^* .
\end{equation}
If we keep only such a sequence in the summation
of (\ref{Gxx}) and use Lemma 1,
\begin{eqnarray}
    G(x, x^*; t-R, t) &\ge &
     G(x, x_1; t-R)G(x_1, x_2; t-R+1)\cdots G(x_{R-1}, x_R; t-1)
     \nonumber \\
     &\ge &
     \prod_{k=1}^R 
       w \left( 1+\frac{(q-1)L}{T(t-R+k-1)}\right)^{1/(1-q)}
     \nonumber \\
     &\ge &
       w^R \left( 1+\frac{(q-1)L}{T(t-1)}\right)^{R/(1-q)} .
\end{eqnarray}
Then the coefficient of ergodicity satisfies
\begin{eqnarray}
  \alpha (G(t-R, t)) &=&
    1-\min \{ \sum_{z\in {\cal S}} \min \{ 
      G(x, z; t-R, t), G(y, z; t-R, t)\} | x, y\in {\cal S}\}
    \nonumber\\
    &\le &
      1-\min \{ \min \{ 
      G(x, x^*; t-R, t), G(y, x^*; t-R, t)\} | x, y\in {\cal S}\}
    \nonumber\\
     &\le &
       1-w^R \left( 1+\frac{(q-1)L}{T(t-1)}\right)^{R/(1-q)} .
\end{eqnarray}
We now use the annealing schedule (\ref{schedule}).
There exists a non-negative integer $k_0$ such that
the following inequalities hold for all $k\ge k_0$:
\begin{eqnarray}
  1-\alpha (G(kR-R, kR)) &\ge & w^R  
        \left( 1+\frac{(q-1)L(kR+1)^c}{b}\right)^{R/(1-q)}
     \nonumber\\
     &\ge & w^R  
         \left( \frac{2(q-1)LR^c}{b}
           \left( k+\frac{1}{R}\right)^c \right)^{R/(1-q)} .
     \label{ceferg}
\end{eqnarray}
It is clear from (\ref{ceferg}) that the summation
\begin{equation}
  \sum_{k=0}^\infty (1-\alpha (G(kR-R, kR)))
    =\sum_{k=0}^{k_0-1} (1-\alpha (G(kR-R, kR)))
    + \sum_{k=k_0}^\infty (1-\alpha (G(kR-R, kR)))
\end{equation}
diverges if $c$ satisfies $0<c\le (q-1)/R$.  This proves weak ergodicity
according to Theorem \ref{weak}.
\qed
%
%
%%%%%%%%%%%%%%%%%%%%%%%%%%%%%%%%

\begin{flushleft}
{\it Remark 1. }
The arguments developed in sections 2 and 3 break
down for $q<1$.
For instance, the argument of the outer parentheses
on the right hand side of (\ref{acceptance})
becomes negative for sufficiently small $T$ if $E(y)-E(x)>0$ and
$q<1$.
The acceptance probability is regarded as vanishing in such a case
in numerical calculations \cite{Tsallis,Penna}.
However, it is difficult to modify the present proof
to adopt this convention used in numerical investigations.
Theorem \ref{convergence} anyway does not exclude the possibility that
the present Markov chain is weakly ergodic for $q<1$
or that it is strongly ergodic for arbitrary $q$.
\end{flushleft}
\begin{flushleft}
{\it Remark 2. }
The condition for weak ergodicity given in Theorem
\ref{weak} is similar to the condition of "infinite often in time"
used to show convergence under the generalized generation
probability in continuum space \cite{Szu,Tsallis}.
\end{flushleft}
\begin{flushleft}
{\it Remark 3. }
Theorem \ref{convergence} with the annealing schedule
(\ref{schedule}) does not immediately mean a fast
convergence of the expectation value of the cost function.
We have proved only the convergence in the sense of
weak ergodicity, not a {\it fast} convergence of the
expectation value of the cost function.  See section 6
for detailed discussions on this point.
\end{flushleft}
%
%%%%%%%%%%%%%%%%%%%%%%%%%%%%%%%%%%%%
\section{Case of $q<1$}
%%%%%%%%%%%%%%%%%%%%%%%%%%%%%%%%%%%
%
It is instructive to investigate a simple solvable
model with the parameter $q<1$
because the general analysis in the previous section
excluded this range of $q$ for technical reasons.
The one-dimensional model discussed by Shinomoto
and Kabashima \cite{Shino} is particularly suited for this purpose.

They considered the thermal diffusion process of an object
in a one-dimensional space.
The object is located on one of the discrete positions
$x=ai$, with $i$ an integer, and is under the potential $E(x)=x^2/2$.
Hoppings to neighbouring positions $i+1$ and $i-1$
take place if thermal fluctuations allow the object to
climb over the barriers with height $B$ for the process
$i\to i-1$ and height $B+\Delta_i$ for
$i\to i+1$,
where $\Delta_i$ is the difference of the potentials at
neighbouring locations
$
  \Delta_i = E\left( a(i+1)\right) -E(ai) .
$
By adaptively optimizing the temperature at each give time,
they found that the energy
(the expectation value of the potential at the position of
the object)
decreases as $B/\log t$.
The optimum annealing schedule $T_{\rm opt}(t)$
was shown to have this same asymptotic
behaviour as a function of $t$.
We show in the present section that the generalized transition
probability with $q=1/2$ leads to a much faster convergence
of the energy.

It should be noted that the analysis of the present section
is not an application of the general theorem in the
previous section.
For example, $q$ is less than 1 here, the number of possible states
is not finite ($i$ runs from $-\infty$ to $\infty$),
and the annealing schedule will turn out
to be $t^{-1}$, not $t^{-c}$.
The purpose of the present section is to show the existence
of a case, independently of Theorem \ref{convergence},
where the generalized transition probability yields
much faster decrease of the temperature and energy.

The problem is defined by the master equation describing
the time evolution of the probability $P_i$
that the object is at the $i$th position at time $t$:
\begin{eqnarray}
  \frac{d P_i}{dt} &=&
    \left( 1+(q-1)\frac{B}{T}\right)^{1/(1-q)} P_{i+1}
    +\left( 1+(q-1)\frac{B+\Delta_{i-1}}{T}\right)^{1/(1-q)} P_{i-1}
       \nonumber\\
    &-&\left( 1+(q-1)\frac{B+\Delta_{i}}{T}\right)^{1/(1-q)} P_{i}
    -\left( 1+(q-1)\frac{B}{T}\right)^{1/(1-q)} P_{i}  .
    \label{master}
\end{eqnarray}
It is straightforward to show that this master equation
reduces to the following Fokker-Planck equation in the continuum
limit $a\to 0$
\begin{equation}
  \frac{\partial P}{\partial t} = \gamma (T)\frac{\partial}{\partial x} (x P)
     +D(T)\frac{\partial^2 P}{\partial x^2}
     \label{FP}
\end{equation}
where
\begin{eqnarray}
  \gamma (T)&=& \frac{1}{T}\left( 1+(q-1)\frac{B}{T}\right)^{q/(1-q)}
    \label{gamma} \\
  D(T) &=&  \left( 1+(q-1)\frac{B}{T}\right)^{1/(1-q)}.
    \label{D}
\end{eqnarray}
We have rescaled the time unit by $1/a^2$ as in \cite{Shino}.

Our aim is to find the fastest possible asymptotic decrease
of the expectation value of the potential defined by
\begin{equation}
  y = \int dx E(x) P(x, t) 
  \label{ydef}
\end{equation}
by adaptively changing $T$ as a function of time.
Differentiating both sides of the definition 
(\ref{ydef}) and using
the Fokker-Planck equation (\ref{FP}), we obtain the
following equation describing the time evolution of $y$:
\begin{equation}
  \frac{dy}{dt} = -2\gamma (T) y + D(T) .
  \label{evol}
\end{equation}
The temperature is adaptively optimized by extremizing this
right hand side with respect to $T$, yielding
\begin{eqnarray}
  T_{\rm opt}&=&\frac{2yB+(1-q)B^2}{2y+B}
     \nonumber\\
   &=& (1-q)B+2qy +O(y^2).
   \label{Topt}
\end{eqnarray}
The evolution equation (\ref{evol}) then has the asymptotic form
\begin{equation}
  \frac{dy}{dt} = -\frac{2}{B} 
  \left(\frac{2q}{1-q}\right)^{q/(1-q)} y^{1/(1-q)} .
\end{equation}
The solution is
\begin{equation}
  y=B^{q/(1-q)} \left( \frac{1-q}{2q}\right)^{1/q}
    t^{-(1-q)/q} .
  \label{ysol}
\end{equation}
The optimum annealing schedule (\ref{Topt}) is now
\begin{equation}
  T_{\rm opt} \sim (1-q) B+ {\rm const} \cdot t^{-(1-q)/q} .
  \label{Topt2}
\end{equation}
The asymptotic behaviour of the
average position can be calculated in the
same way.  The result is
\begin{eqnarray}
  \langle x \rangle &=& \int dx \, x P(x, t) \nonumber \\
    &\sim&  {\rm const}\cdot t^{-1/2q} .
\end{eqnarray}

It is useful to restrict of the value of $q$
to avoid unphysical behaviour of the generalized transition
probability in the present one-dimensional problem.
One of the transition probabilities in the master equation
(\ref{master})
\begin{equation}
  \left( 1+(q-1)\frac{B+\Delta _{i-1}}{T}\right)^{1/(1-q)}
\end{equation}
reduces for $T=(1-q)B+O(y)$ to
\begin{equation}
  \left( \frac{a^2-2ax}{2B} +O(y) \right)^{1/(1-q) } .
\end{equation}
This quantity must be a small positive number for any $x$
and sufficiently small (but fixed) $a$.
This requirement is satisfied if
\begin{equation}
  q=1-\frac{1}{2n} \quad (n=1,2,\cdots) .
  \label{qs}
\end{equation}
Consistency of the other transition probabilities in (\ref{master})
is also guaranteed under (\ref{qs}).

From (\ref{ysol}) we see that the fastest decrease of the energy
is achieved when $q=1/2$.  With this value of $q$,
\begin{eqnarray}
  y &\sim & \frac{B}{4} \, t^{-1}  
    \label{yopt}\\
  T_{\rm opt} &\sim & \frac{B}{2} +\frac{B}{4} \, t^{-1}
     \label{Topt3}\\
  \langle x \rangle  &\sim &{\rm const}\cdot t^{-1} .
\end{eqnarray}

It may be useful to remark that
the non-vanishing value $(1-q)B$ of the temperature
(\ref{Topt2}) in the infinite-time limit does not cause troubles.
What is required is not an asymptotically vanishing value of the
temperature but that the probability distribution
does not change with time in the infinite-time limit.
This condition is satisfied if $T=(1-q)B$
as is apparent from
(\ref{FP}) with (\ref{gamma}) and (\ref{D}).

The results (\ref{yopt}) and (\ref{Topt3})
show asymptotic relaxations proportional to $t^{-1}$ which is
much faster than those for the conventional
transition probability, $B/\log t$ \cite{Shino}.
This result of course depends upon the specific structure
of the one-dimensional model.
We are not claiming to have shown
that the generalized transition
probability with $q<1$ always gives faster
decrease of the temperature and energy.
%

%%%%%%%%%%%%%%%%%%%%%%%%%%%%%%
\section{More general transition probability}
%%%%%%%%%%%%%%%%%%%%%%%%%%%%%%
%
A natural question may arise on how far the arguments
in the previous sections depend on the specific form of the
acceptance probability (\ref{acceptance}).
We investigate this problem for the one-dimensional
model treated in the preceding section.

The master equation is now generalized to
\begin{equation}
  \frac{d P_i}{dt} =f\left(\frac{B}{T}\right) P_{i+1}
    +f\left(\frac{B+\Delta_{i-1}}{T}\right) P_{i-1}
    -f\left(\frac{B+\Delta_{i}}{T}\right) P_{i}
    -f\left(\frac{B}{T}\right) P_{i}  .
    \label{gmaster}
\end{equation}
The same Fokker-Planck equation (\ref{FP}) is derived
in the limit $a\to 0$ with the following parameters
\begin{eqnarray}
   \gamma (T)&=&-\frac{1}{T^2} f'\left(\frac{B}{T}\right)
    \\
   D(T)&=& f\left(\frac{B}{T}\right) .
\end{eqnarray}
The expectation value of the potential obeys the same
evolution equation as in (\ref{evol}):
\begin{equation}
  \frac{dy}{dt} = -2\gamma (T) y + D(T) 
    =\frac{2y}{T}f'\left(\frac{B}{T}\right) +f\left(\frac{B}{T}\right)
    \equiv {\cal L}\left( \frac{1}{T}\right) .
  \label{evol2}
\end{equation}
Minimization of ${\cal L}(v) (v=1/T)$ with respect to $T$ for
given $y$ leads to
\begin{equation}
  2vyB f''(Bv) + (2y+B) f'(Bv) =0 .
   \label{Lp}
\end{equation}
The solution of this equation for $v$ gives the optimal annealing
schedule
\begin{equation}
  \frac{1}{T_{\rm opt}} = v = g(y) .
    \label{Toptg}
\end{equation}
Assuming analyticity of $g(y)$ as $y\to 0$,
we write (\ref{Toptg}) as
\begin{equation}
  v=c_1 +c_2 y +O(y^2) .
    \label{vsol}
\end{equation}

It is required that the system stops its time evolution as
$y\to 0$ and $v\to c_1$.
We then have $f(B c_1)=0$ from
(\ref{evol2}) assuming $c_1$ is finite.
(This condition of $c_1<\infty$ is not satisfied by the
conventional Boltzmann-type acceptance probability
in which $1/v=T\to 0 (c_1\to\infty)$ as
$y\to 0$.)
It is also necessary that the minimization
condition (\ref{Lp}) is satisfied in the same limit, leading
to $f'(B c_1)=0$.
These two conditions on $f$ and $f'$ are satisfied if
$f(Bv) (=f(Bc_1+Bc_2 y))$ and its derivative behave for small $y$ as
\begin{equation}
 f(Bv) \sim c_3 y^k,\quad f'(Bv)\sim -c_4 y^{k-1} ,
   \label{f4}
\end{equation}
where $k>1$ and $c_3, c_4>0$.  The minus sign in front of
$c_4$ comes from the observation that an increase of the inverse
temperature $v=1/T$ means a decrease of the energy
$y$ and therefore the differentiations by $v$ and $y$
should be done with the opposite sign ({\it i.e.} $c_2<0$).

The evolution equation (\ref{evol2}) then has a form
\begin{equation}
   \frac{dy}{dt} = -c_5 y^k 
\end{equation}
with positive $c_5$ if $2c_1 c_4>c_3$.  This equation is solved as
\begin{equation}
  y=\left( c_5 (k-1) t\right)^{-1/(k-1)} ,
  \label{ysol2}
\end{equation}
which shows a power decay of the expectation value
of the cost function.

It is useful to set a restriction on $k$ as in
the preceding section for $q$.
The following acceptance probability for $y\to 0$
should be positive for any $x$:
\begin{equation}
f\left(\frac{B+\Delta_{i-1}}{T}\right)
 \sim f(B c_1 + B c_1 \Delta_{i-1} )
 \sim \left(\frac{a^2}{2}-ax\right)^k ,
\end{equation}
where we have used (\ref{f4}).
This requirement is satisfied if $k$ is a positive even number $k=2n$.
The energy (\ref{ysol2}) then decays as
\begin{equation}
  y \sim t^{-1}, t^{-1/3}, t^{-1/5},\cdots ,
\end{equation}
the same formula as in the preceding section.
In fact the argument in section 4 is recovered if we choose
\begin{equation}
  f(v)= \left( 1+(q-1) v \right)^{1/(1-q)} .
\end{equation}

In this way the fast decrease of the energy has been obtained
for a very general acceptance probability distribution function
satisfying certain analyticity conditions.
%

%%%%%%%%%%%%%%%%%%%%%%%%%%%%%%
\section{Discussions}
%%%%%%%%%%%%%%%%%%%%%%%%%%%%%%
We have proved weak ergodicity of the
inhomogeneous Markov process generated by
the generalized transition probability under certain conditions
on the parameters.
For technical reasons we were unable to prove strong ergodicity,
or more strongly, convergence to the optimal distribution function.
We could not show that the condition (\ref{scond})
of Theorem \ref{strong} is
satisfied by the present inhomogenous Markov chain.
However, weak ergodicity alone already means that the state of the system
asymptotically becomes independent of the initial condition,
and it is most likely that such an asymptotic state is the optimal one as
mentioned in section 1.

It is appropriate to comment on computational complexity
here. 
The time $t_1$ necessary for the temperature (\ref{schedule})
to reach a small specified value $\delta$ is obtained by
solving the relation
$b/t_1^c \sim \delta$ ($c=(q-1)/R$) for $t_1$:
\begin{equation}
  t_1 \sim \exp \left( \frac{k_1 N}{q-1}\log \frac{b}{\delta}\right) .
  \label{t1}
\end{equation}
Here we have set $R=k_1 N$ with $N$ the system size
because $R$ defined in (\ref{R}) is roughly of this order
of magnitude in many cases.  For example, in the problem of
spin glasses, one can reach any spin configuration by flipping
at most $N$ spins.
The corresponding time for the conventional simulated annealing is
\begin{equation}
  t _2\sim \exp \left( \frac{k_2 N}{\delta} \right)
  \label{t2}
\end{equation}
which has been obtained from $k_2 N/\log t_2 \sim \delta$.
A comparison of (\ref{t1}) with (\ref{t2}) reveals that the
coefficient of $N$ in the exponent has been reduced from
$1/\delta$ to $\log 1/\delta$ by using the generalized transition
probability.
In this sense, $t_1 \ll t_2$.
Since we have proved Theorem \ref{convergence} under very
general conditions on the system (which would include
problems with NP completeness), it is not possible
to find an algorithm to reach a low-temperature state
in polynomial time.
The best we could achieve is an improvement of the
coefficient in the exponent.

One should be careful that the rapid decrease of the temperature
does not immediately mean a rapid decrease of the cost function.
This aspect can be checked by comparing the
acceptance probability (\ref{acceptance}) at $T=\delta$
\begin{equation}
 u_1(T=\delta ) \sim \left(\frac{\delta}{(q-1)\Delta E}\right)^{1/(q-1)} 
  \label{prob1}
\end{equation}
with the corresponding one for the conventional transition
\begin{equation}
  u_2(T=\delta ) \sim \exp (-\Delta E/\delta ) .
   \label{prob2}
\end{equation}
Since $u_1(\delta)\gg u_2(\delta)$ if $\Delta E/\delta \gg 1$,
we see that the generalized transition probability
at a given temperature has a larger value to induce
transitions into states with
high values of the cost function than in the case of the
conventional one at the same temperature.
Thus the expectation value of the cost function may
be larger under the generalized
transition probability than under the conventional
Boltzmann form at the same temperature if one waits
sufficiently long until thermal equilibrium is reached.
This phenomenon has actually been observed in a numerical
investigation under a slightly different (but essentially
similar) situation \cite{Andre}.

Therefore, if the expectation value of the cost function
is observed in a numerical simulation
to indeed decrease rapidly under the generalized
transition probability, it would be not only for the
rapid decrease of the temperature but also because
the relaxation time is shorter.
The conventional transition probability may give a larger
possibility for the system to stay longer in local minima
with high values of the cost function.
A mathematical analysis of this property of
quick relaxation by the generalized transition probability
is beyond the scope of the present paper.
However one may naively expect it to happen
from the larger probability to climb over high
barriers as discussed above.

It should be remarked that Theorem \ref{convergence} with
the annealing schedule (\ref{schedule}) does not
give a practically useful prescription of
simulated annealing.
In actual numerical simulations one rarely uses
such annealing schedules as (\ref{schedule})
obtained from worst-case estimates.
Even exponentially fast decreases of the temperature
often give satisfactory results 
in the conventional and generalized methods
(see \cite{Penna} and references in \cite{Tsallis}).
The significance of Theorem \ref{convergence}
is that convergence
(in the sense of weak ergodicity) has anyhow been 
proved with the annealing schedule (\ref{schedule})
under the generalized transition (acceptance) probability
where only empirical numerical investigations have been
carried out without mathematical guarantee of convergence
under any annealing schedule.

\vspace{3mm}
We thank Dr Naoki Kawashima, Dr Toshiyuki Tanaka, Dr Tsuyoshi Chawanya,
and Prof Constantino Tsallis for discussions and comments.
One of the authors (J. I.) acknowledges support from the
Junior Research Associate Program of RIKEN.

%%%%%%%%%%%%%%%%%%%%%%%

\end{document}